\documentclass[twocolumn]{aa}
\usepackage{graphicx}
\usepackage{natbib}
\bibpunct{(}{)}{;}{a}{}{,}
\begin{document}
\title{Gamma-Ray All-Sky Imaging with the Burst and Transient Source Experiment}

\titlerunning{Gamma-Ray All-Sky Imaging with BATSE}

\author{S.E. Shaw\inst{1}
\thanks{\emph{Present address:} INTEGRAL Science Data Centre, Chemin d'Ecogia 16, CH-1290 Versoix, Switzerland}
\and M.J. Westmore\inst{1}
\and A.B. Hill\inst{1}
\and A.J. Bird\inst{1}
\and A.J. Dean\inst{1}
\and C. Ferguson\inst{1}
\and \\ J. Kn\"{o}dlseder\inst{2}
\and J.J. Lockley\inst{1}
\thanks{\emph{Present address:} Oxford University Computing Centre, Wolfson Building, Parks Road, Oxford, OX1 3QD, UK}
\and D.R. Willis\inst{1}
%\and E. J. Barlow\inst{1}
%\and T.V. Tikkanen\inst{1}
}
\institute{School of Physics and Astronomy, University of Southampton, SO17 1BJ, UK
\and Centre d'Etude Spatiale des Rayonnements, 9, avenue Colonel-Roche, B.P. 4346, 31028 Toulouse Cedex 4, FRANCE}

\offprints{S.E. Shaw, \email{simon.shaw@obs.unige.ch}}

\date{Accepted for publication in Astronomy and Astrophysics February 2004.\\{\em This is the astro-ph version, with lower resolution figures than in the A\&A publication.}}

\abstract{
The BATSE mission aboard CGRO monitored the whole sky in the 20 keV - 1 MeV energy band continuously from April 1991 until June 2000.  Although BATSE had very poor intrinsic angular resolution, the data can be used to survey the entire soft gamma-ray sky with $<1^{\circ}$ angular resolution and $\sim$mCrab sensitivity by using the Earth occultation method.  This method determines flux by measuring the step in the count rate profile in each BATSE detector as a source rises above or sets below the Earth's limb.  A maximum likelihood imaging technique can then be used to build up all-sky maps from the images of the Earth's limb produced by occulting sources.  However, since the Earth seen from BATSE has a radius of $\sim$70$^{\circ}$, the limb images that intersect at the positions of bright point sources have a significant effect over the area of the all-sky map.  A method for performing image cleaning on likelihood data has also been developed and is used to effectively remove artefacts from the all-sky maps.  This paper describes the 'LIMBO' imaging technique and presents preliminary all-sky maps of 25 - 160 keV emission, the first to be made since the HEAO1-A4 mission of 1978-79.

   }

\maketitle

%
%________________________________________________________________

\section{Introduction}
Conventional imaging techniques, of focusing light with reflecting or refracting surfaces, are not possible in the hard X-ray or gamma-ray regions of the electromagnetic spectrum due to the highly penetrating nature of these photons (however, see \citet{ballmoos} for a report on the first use of a Bragg diffraction gamma-ray lens in astrophysics).  For photons above $\sim 10$ keV, where grazing incidence techniques become highly inefficient, it is necessary to make images by transforming the spatial information of the object onto the detector plane with some sort of modulation or coding technique that can be used to form a reconstructed image.

The ability to form an image has many advantages over simple measurements of source fluxes made with pointings of basic collimated instruments.  The ability to locate a source with an image allows comparison with measurements made at different times, with different instruments at different wavelengths.  Crowded fields can be resolved, minimising the possibilty of source confusion and the nature of extended or diffuse emission, such as that which may be seen from Supernova Remnants (SNR) or the Galactic Plane (GP), can be investigated.  Imaging detectors can also have a large gain in sensitivity over simpler instruments by reducing the amount of time needed to make accurate observations;  It is possible to make an estimation of the background in the region of a source at the same time as the observation, negating the need to perform control observations away from the source.  In an instrument with a large Field of View (FOV) it is possible to observe many sources in a single observation, which is often referred to as the \emph{multiplex advantage}. 

\begin{figure*}[ht]
\centering
\includegraphics[width=17cm]{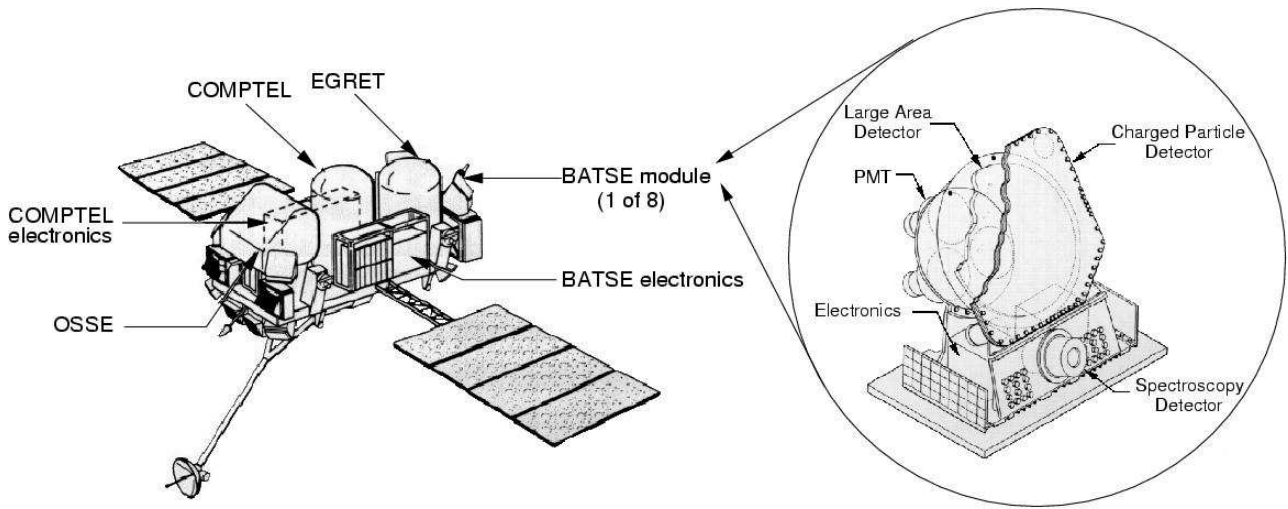}
\caption{The Compton Gamma-Ray Observatory and one of the 8 BATSE detector modules (drawings courtesy of the National Space Science and Technology Centre, Huntsville, AL, USA).}
\label{fig:cgro}
%\vspace{150mm}
\end{figure*}

The natural progression of the last point above is to image the entire celestial sphere and perform an all-sky survey.  Knowing the locations and properties of a large number of sources is important in all energy bands of astronomy if only for the purposes of design, planning and proposal justification of other instruments and observations.  Even a survey with crude angular or spectral resolution can be highly important; as long as the sources in the survey catalogue have well determined uncertainties then other observations by higher resolution instruments are possible.  Contemporaneous measurements of a large number of sources are important in conducting morphological studies of families of objects (e.g. \citet{seyferts}).  This is possible for an instrument with a large FOV, which is able to conduct a survey of such sources free from the biases that may be introduced by conducting many observations on different smaller FOV instruments at different times.  Similarly, by observing a large part of the sky, it is possible to maximise the probabilty of detecting unknown or transient objects, such as Gamma-Ray Bursts (GRBs) or X-ray novae, in a fixed time.  The nature of hard X-ray and gamma-ray imaging systems means that they often have much larger FOVs than is possible at optical wavelengths.  In the extreme case it is possible to observe the entire celestial sphere at once.  

In high energy astronomy all-sky images have been provided between 0.1 - 2 keV by ROSAT \citep{rosat} and $>$ 1 MeV by COMPTEL \citep{comptelal26, comptelcat}, but there is sparse information in the intervening band, provided almost exclusively by the HEAO1-A4 experiment of 1978-79 \citep{HEAO1A4}.  This experiment consited of two NaI/CsI phoswich detectors arranged so that they observed simultaneously in opposite directions and spun about an axis perpendicular to the optic axis and in the plane of the ecliptic.  This meant that it took 6 months for the whole sky to be observed and so most areas of the sky were only sampled two or three times during the 16 month mission.  The HEAO1-A4 survey catalogue contains 72 sources in four broad energy channels ranging from 13 - 180 keV, at a flux sensitivity of 14 mCrab.    The A4 experiment suffered from some problems with changing efficiencies between the two detectors and the sensitivity in the high energy band was limited; only 14 of the sources were detected at energies greater than $80$ keV.  Although the HEAO1-A4 was an important mission the usefulness of the survey is limited by the low spectral range and resolution and the fact that many transient sources were missed because of the slow sampling of the whole sky.

The aim of this work is to show how the wealth of the BATSE data set, which comprises 9 uninterrupted years of all-sky monitoring, can be exploited to achieve the maximum scientific gain.    

\section{The Burst and Transient Source Experiment}
\label{sec:batse}
The Burst and Transient Source Experiment (BATSE) monitored the whole
sky continuously from April 1991 - June 2000 as an experiment aboard
the Compton Gamma-Ray Observatory (CGRO) \citep{batse:fish89a}.  It
was designed primarily to detect and locate GRBs
\citep{batse:fish89b}.  BATSE consisted of eight uncollimated 2025
cm$^{2}$ NaI(Tl) Large Area Detectors (LADs), which had a combined
field of view of $4\pi$ steradians and were sensitive from 20 keV - 1
MeV.  Also included in each detector module was a spectral detector
(SD) made of a smaller 127 cm$^{2}$ NaI(Tl) crystal.  The SDs were
optimized for energy coverage and resolution, and were operated over
the range 10 keV - 100 MeV.  An illustration of CGRO and a BATSE
module can be seen in Fig.~\ref{fig:cgro}.

Data from the detectors were recorded in several formats,
  of which two recorded by the LADs are of primary interest for this work: the DISCLA data, consisting of 4 energy
channels recorded in 1.024 second bins and CONT data, comprising 16
energy channels with a 2.048 second resolution.  Triangulation between
LADs was used to locate GRBs crudely, to within a few degrees,
depending on the strength of the flux \citep{briggs99, pend99}.  BATSE was a highly succesful experiment that detected almost 3000 GRBs in a mission which far exceeded the nominal lifetime of the experiment.  The experiment was still providing good quality data when concerns over degradation in the performance of the CGRO attitude control system lead to the satellite being de-orbited safely over the Pacific in June 2000.  The identification of GRBs with cosmological rather than local events can be largely attributed to the wealth of data collected by BATSE \citep{batse:meegan92}.  

\section{Earth Occultation Imaging with BATSE}
\label{sec:eot}
It was realised before the CGRO launch that the angular resolution for
non-transient gamma-ray  sources could be greatly improved, over that
obtainable for GRBs, with knowledge of the position on the sky of the
Earth's limb \citep{fishman82,paciesas85,batse:fish89b}.  A short
description of Earth occultation analysis of persistant gamma-ray
sources is given below, but the reader is directed to the work of
\citet{batse_eot} for a full discussion of the technique.

\begin{figure}
\resizebox{\hsize}{!}{\includegraphics{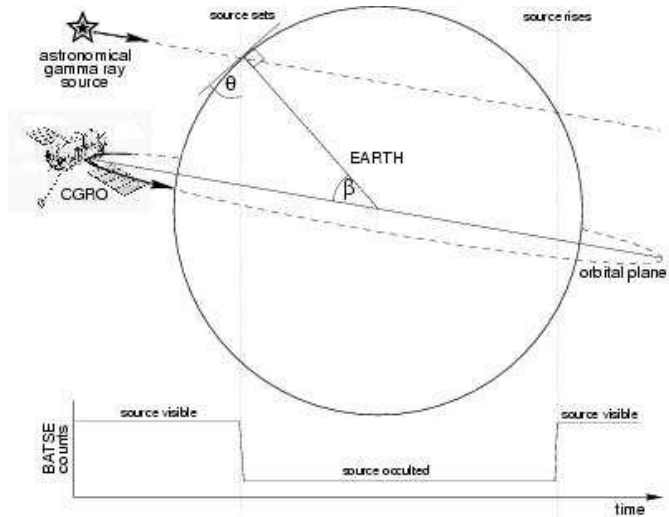}}
\caption{A schematic illustrating the Earth Occultation technique for detecting astronomical gamma-ray sources with BATSE.  During the CGRO spacecraft's orbit, measurable steps are seen in the BATSE detector count rates as a gamma-ray emitting object sets below and rises above the Earth's horizon.  The occultation angle, $\beta$, is defined as the angle between the orbital plane and the object measured at the geocentre. }
\label{fig:simpleocc}
%\vspace{150mm}
\end{figure}

The Earth Occultation technique is a way of performing imaging, with
simple non-pixellated or low spatial resolution detectors, by using
the Earth as a temporal modulator of the gamma-ray flux.  The
technique is possible with the BATSE SDs (see, for example,
\cite{mcnamara}) however, due to their larger size, and hence
sensitivity, data from the LADs only has been used in this work.  The Earth,
seen from a BATSE detector, has a diameter of $\sim$140$^{\circ}$ and
will sweep rising and setting arcs across the sky at the CGRO orbital 
period, $\sim$90 minutes.  As gamma-ray emitting objects set below the
Earth's horizon downward steps are seen in the detector count rates at
the orbital period (and vice-versa for rising objects, see Fig.~\ref{fig:simpleocc}).  The height above the background of the steps
gives a direct measurement of the flux of the source being occulted.
With knowledge of the detector orientation and the relative position
in the orbit of the spacecraft with respect to the Earth, it is
possible to project an image of the Earth's limb on the sky at the
time of the source occultation.  An image of the gamma-ray emission
from the sky can then be constructed by building up the images of the
Earth's limb from many occultations.  The intrinsic angular resolution
depends on the rise time of an occultation step.  For a source
crossing Earth in the CGRO orbital plane the transit time from full
transmission to full attenuation, $\Delta t$,  is $\sim$8 seconds, due
to the increasing absorption of gamma-rays in the varying column
density of the atmosphere as the source decreases in altitude.  As the
source occultation happens further away from the orbital plane $\Delta
t$ increases as $1/\cos{\beta}$, where $\beta$ is the occultation
angle measured at the geocentre between the orbital plane and the
source. $\Delta t$, for a source seen by a satellite with orbital
period $P$ translates to an angle of $360^{\circ}\times \Delta
t/P$. This angle represents the optimum angular resolution, for
imaging a source, that is obtainable with this technique.  Hence, for
the BATSE/CGRO system the ideal angular resolution is $\sim
0.5^{\circ}$.  When $\beta > 70^{\circ}$ no occultation is seen
although the occultation is effectively unusable when $\beta >
68^{\circ}$ as the source skims across the poles of the Earth \citep{batse_eot}.

The inclination of the CGRO orbit was 28$^{\circ}$ with a precession period of 53 days.  This orbit is such that the whole sky undergoes occultation during the precession period, although regions of sky near the poles with $\vert 41^{\circ}<\delta < 82^{\circ}\vert$ receive less coverage than the rest of the sky.  In total, including effects for lost data etc., between 50\% and 90\% of the entire sky can be observed in any 53 day period.  The Earth occultation technique, when used with an instrument such as BATSE with a large field of view is a very attractive tool for the production of all-sky images.

\subsection{Standard Occultation Analysis Techniques and their Limitations}
\label{sec:others}
Earth occultation techniques have been applied to the BATSE database
since the launch of CGRO to observe the gamma-ray sky.  Techniques
were developed by workers at the Marshall Space Flight Centre (MSFC, note that the science groups of MSFC were moved to the
  National Space Science and Technology Centre (NSSTC) in 2001) to
allow fast monitoring of bright transient or flaring sources by
searching through the count rates for occultation step-like features
(see \citet{batse_eot} for a full explanantion).  By measuring the
occultation time the location of the source could be deduced, from the
orientation of the Earth's rising and setting limbs, and a position
determined to $\sim 1^{\circ}$ accuracy.  Examples of the use of this
technique are the discovery of the gamma-ray sources GRS 1716-249
\citep{grs1716} and GRO 0422+32 \citep {gro0422} and observations of
the X-ray nova 4U 1543-47 \citep{4u1543}.  However, the sensitivity of
the step searching technique is limited by the difficulty in
accurately determining the background over several orbits and by
non-statistical noise introduced by bright pulsating sources with
periods of order a few minutes, such as Vela X-1.  ~~\citet{batse_eot}
quote a 3$\sigma$ sensitivity of $\sim 200$ mCrab for a source with a spectrum like that of the Crab Nebula  ($4.6\times 10^{-9}$ ergs cm$^{-2}$s$^{-1}$ for 30 - 250 keV).  

For known sources a fitting procedure can be used, on a 20 second
fitting window, applied within a short (220 or
240 second) section of data, centred on the
expected occultation of the source in question.  For each LAD and
energy channel the model fit consists of a quadratic curve describing
the background and a set of response vectors that model the effects of
all sources contained within the fit window (i.e. the target source
and any other interfering background source).  This allows multiple
steps to be combined, giving an improvement in sensitivity, over
the step search technique, proportional to the square root of the
number of steps used.  In one day it is possible to have between 0 and
32 steps, and so the the improvement in sensitivity is about a factor
of four (since the average number of steps used is 16).  In principle increasing the size of the fit window reduces the statistical uncertainty in measuring the step size and hence improves the sensitivity.  However, the length of fit window is constrained heavily by the ability of the background model to describe variations in the count rate.  As the fit window increases a more complex model may be needed to account for slower varying terms in the background as well as accounting for the higher probabilty of encountering other steps from faint, unknown or flaring sources.  Indeed, the effects of interfering sources could account for a 30-50\% error in determining BATSE fluxes \citep{mattsthesis, malizia99}.  Hence a compromise must be made to reduce systematic uncertainties in the fit at the expense of increased statistical error.  

The step searching and fitting techniques provide a flux and a position but are not capable of producing images.  In particular the only consistent way of dealing with the problems of interfering sources in the fit window is to develop a generalised, unbiased imaging capabilty.  The timing information about occultation steps in the count rate light curve can be transformed into spatial information for the construction of images.  The time derivative of the count rate in a BATSE detector is equal to the line integral of the flux along the Earth's limb.  During an occultation the flux is sampled at discrete time intervals as the occulting part of the Earth's limb tracks across the source at a constant angle, $\theta(\beta)$ (see Fig.~\ref{fig:simpleocc}).  For a single location on the sky, the flux will be sampled at different angles for subsequent rises and sets of the source (except in the case where $\beta = 0$) and as the CGRO orbit precesses, allowing the reconstruction of the image.  The occultation transform imaging technique uses a combination of Radon transformation and maximum entropy methods to make images from BATSE occultation data \citep{zhang93nature,zhang95expA}.  The technique has proved successful in locating several new transient sources with better than $0.5^{\circ}$ accuracy \citep{grs1009,gro1655} and also allowed the observations of weak emission from low mass X-ray binaries, active galactic nuceli and supernova remnants \citep{barret,malizia,mccolloughsnr}.  

\begin{figure*}
\includegraphics[width=17cm]{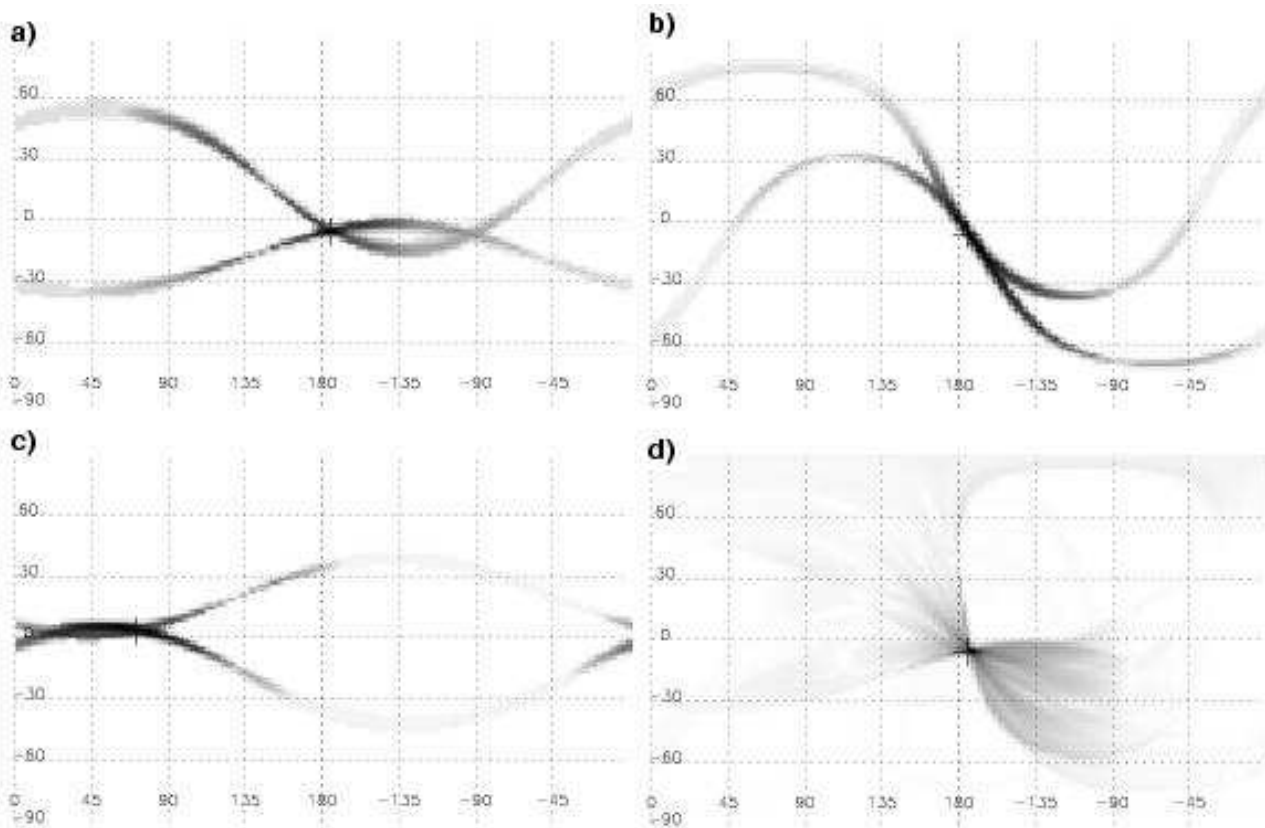}
\caption{The projections on the sky, in galactic coordinates, of
  occultation limbs for sources seen by BATSE during a 24 hour period:
  a) The Crab Nebula for TJD 09586; b) The Crab Nebula for TJD 09560;
  c) Cygnus X-1 for TJD 09586.  The time difference between TJD 09560 and TJD 09586 is approximately 50\% of the CGRO orbital precession period.  Note how the quality of positional information in two dimensions varies with source position and time.  The result of adding limbs for a full 53 day precession
  cycle is shown in d) and shows how the Crab Nebula position is well
  defined with no positional ambiguity caused by secondary intersections of the limbs, such as is seen in a) and c).  However, it is clear that the angular sampling of the source is still not complete.  The crosses mark the positions of the sources.}
\label{fig:crabpsf50}
%\vspace{150mm}
\end{figure*}

The occultation transform imaging technique suffers from a number of problems that limit its usefulness for observing weak sources and crowded regions, and  make it ineffective for the production of all-sky images.  Firstly the use of iterative non-linear reconstuction techniques, such as the maximum entropy method are significantly computationally intensive.  This makes the consideration of long sections of data impractical and it is unclear how images constructed from a number of short sections can be combined \citep{mattsthesis}.  Also, the Radon transformation uses an implicit assumption that the Earth's limb is linear and so only a small part of the occulting arc can be considered.  This effectively limits the field of view of the final image to 20$^{\circ}\times$20$^{\circ}$, since larger images produce distortion of point source locations near the edges of the field.  The Radon transformation, as applied to the occultation transformation, states that an image $M(x,y)$ can be reconstructed from a set of $g(\theta, t)$ corresponding to the time series of the parallel line integrals, inclined at an angle $\theta$ and swept across $M(x,y)$ in time $t$.  In principle, a simple point source can be reconstructed from two orthogonal $\theta$.  However, in practice, the object may not be a point source and there is likely to be more than one source in the field of view.  The effects of systematic and random noise mean that $\theta$ must be well sampled in order to accurately reconstruct the image.  The sampling of $t$ is also incomplete if the field of view of the reconstructed image is restricted.  A bright source just outside of the reconstructed field of view will make a contribution to the background flux in the image.  Fig.~\ref{fig:crabpsf50}d shows the projection on the sky of images of the Earth's limb seen by BATSE during occultations of the Crab Nebula, built up over a 53 day period.  It can be clearly seen that a source has an influence over a very large part of the sky image and also shows that, even over a whole precession cycle of the CGRO orbit, $\theta$ is not completely sampled.

\section{All-Sky Imaging with Maximum Likelihood}
To fully exploit the $4\pi$ FOV of the BATSE instrument it is necessary to take a different approach in reconstructing the image to the deconvolution used in the occultation transform method, outlined in Sec.~\ref{sec:others}, which is limited to small areas of the sky.  The Likelihood Imaging Method for BATSE Occultation Data (LIMBO) code is a collection of computer programmes developed at Southampton to create all-sky maps from the BATSE data and is based on initial work by J. Kn\"{o}dlseder of Centre d'Etude Spatiale des Rayonnements (CESR) \citep{jurgen,shaw:alicante}.  A pixellated image is built up by using a Maximum Likelihood Ratio (MLR) test for source emission from a given source location on a predefined grid of sky positions.  The problem of image contamination by sources outside the image FOV, seen in the occultation transform technique is avoided by this method as the total dataset, comprising the whole sky image, is used in the calculation of the MLR for each pixel.  It should be stressed that this technique differs from the occultation transform method in that it produces maps of the results of a number of statistical tests for emission rather than a deconvolution of the data.  There are two main parts to the LIMBO imaging process, the background corrected data is first transformed through a differential filter before the likelihood tests are applied and the maps created. 

\subsection{Flat Fielded Data}
\label{sec:bamm}
When considering long sections of BATSE data a good understanding of the highly variable gamma-ray background encountered by the instrument in orbit is essential to enable the reduction of systematic errors introduced in flux determination.  \citet{jpl2000} of the Jet Propulsion Laboratory (JPL) group report the use of essentially similar analysis techniques to MSFC, but with a multi-parameter background fit based on a combination of empirical and semi-empirical functions derived, in part, from data taken by the spacecraft.  This allows them to extend the length of data to be fitted to 1 day.  However, \citet{batse_eot} note that there may be limiting systematic uncertainties remaining, which manifest themselves as significant uncharacteristically hard emission being reported in a number of the published JPL source spectra.

\begin{figure*}
\centering
\includegraphics[width=17cm]{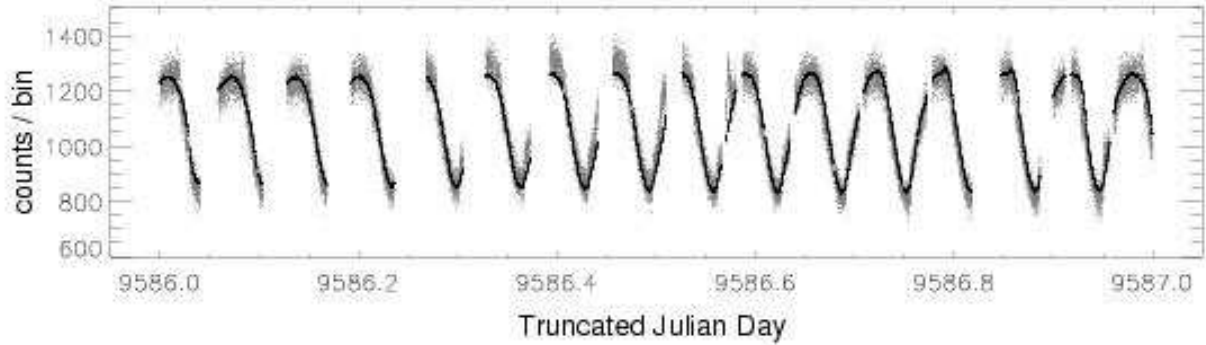}
\caption{The BATSE CONT LAD 1 25 - 35 keV count rate for TJD 9586 (shown in grey) with corresponding results from the BAMM model overlaid (dark curve).  BAMM reproduces the shape, amplitude and variations in the BATSE count rates very accurately: No further fitting or renormalising of the model has been performed in this graph, except for the addition of a constant offset of 351.5 counts per bin to the model for display purposes only.  The counts are shown per 2.048 second CONT time bin.}
\label{fig:bgvbamm}
%\vspace{150mm}
\end{figure*}

\begin{figure*}
\centering
\includegraphics[width=17cm]{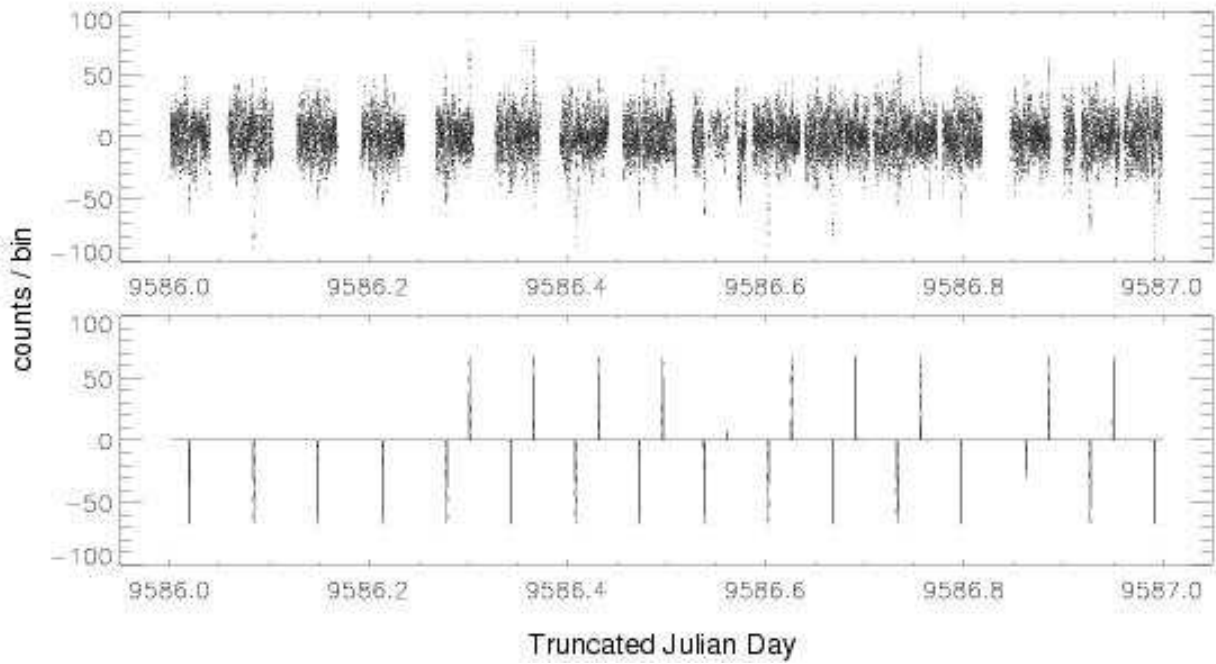}
\caption{Top:  The result of passing the flat-fielded BATSE CONT LAD 1 25 - 35 keV count rate for TJD 09586 through the differential filter of Eq.~\ref{eqn:difffilt}.  The counts show clear peaks, due to the occultations of bright sources.  The peak positions can be compared with the bottom graph, which shows the expected differential response vector multiplied by the value of the flux scaling parameter, $\alpha$, for a source at the Crab Nebula position.  The gaps and shortening of some response peaks are due to missing data in the CONT stream.  In both graphs the filter applied was $f_{a}=4, f_{b}=14$.}
\label{fig:occresp}
%\vspace{150mm}
\end{figure*}

The BATSE Mass Model (BAMM) is a Monte-Carlo based background model
and was developed at Southampton primarily for use with the LIMBO
technique.  Information about the BATSE detector orientations, the
position of CGRO in orbit in relation to the Earth and the local
magnetic and radiation environment are taken from the data stream and
used as the basis for a physical background calculation based on a
database of simulated particle fluxes.  By considering the energy
dependent contributions to the background from three components
(cosmic diffuse and atmospheric albedo flux gamma-rays, prompt cosmic
ray interactions).  The BAMM model is capable of reducing the
variations in the BATSE background flux, by a factor of 8-10, simply
by subtracting the modelled flux from the data.  This effectively
nullifies all systematic effects on the imaging analysis, producing a
`flat-fielded' data set.  A full description of BAMM is outside the scope of this document but is the subject of other works \citep{bammpaper, deanrg2000}.  A general discussion of mass modelling techniques for applications in high energy astrophysics can also be found in \citet{ssr2002}.  An example of the effectiveness of BAMM in accurately modelling the BATSE count rates can be seen in Fig.~\ref{fig:bgvbamm}.

\subsection{Differential Filter}
As CGRO orbits the Earth a path of sky of width up to 140$^{\circ}$  may be occulted by either the rising or setting limb.  Hence it is likely that many sources of varying strengths will be observed, leading to a superposition of steps in the count rate.  The useful information about the source flux is contained in the short section of data surrounding the step itself, therefore it is attractive to deal not with the absolute detector counts but with their time rate of change.

The flat-fielded BATSE data is passed to a simple differential filter, given by
\begin{equation}
o_{i} = \frac{\sum^{j=i+f_{b}}_{j=i+f_{a}}r_{j} - \sum^{j=i-f_{a}}_{j=i-f_{b}}r_{j}}{f_{b}-f_{a}+1}
\label{eqn:difffilt}
\end{equation}
The filtered data vector, $\vec{o}$, is transformed from a count rate,
$\vec{r}$, containing a superposition of occultation steps to a set of
occultation peaks by the filter with inner and outer bounds given by $f_{a}$ and $f_{b}$ respectively (see Fig.~\ref{fig:occresp}).  A square step at bin $i$ will be transfromed to a triangular peak of width $2f$, where $f = f_{b}-f_{a}$.    The choice of $f$ has implications for the sensitivity and angular resolution of the imaging technique; larger $f$ will give a more significant occultation peak and improve sensitivity, smaller $f$ returns less significant but sharper peaks that will lead to better angular resolution.  The filter properties are discussed further in Sect.~\ref{sec:filt}.

The raw BATSE data contains occasional gaps of null data where, for example, the instrument is turned off during passage through the South Atlantic Anomaly (SAA) or where contact is lost between CGRO and its telemetry relay satellites.  Left untreated the gaps cause large artificial peaks in the filtered data near each gap, which mimic sources.  The result of differentiating bin $i$ just before a gap is a large positive peak since Eq.~\ref{eqn:difffilt} depends on counts from bins $i+f$, at least some of which will be null data from within the gap.  One way to reduce this effect would be to ignore all data within $f$ bins of a gap, although this may remove $\sim$ 10\% from the total of good data available.  Instead a simple linear interpolation is used to pad the gaps in the flat-fielded data, this stops the filter blowing up in gap-data boundaries and conserves the maximum of available data.  The padded data from within the gap is marked as bad in the filtered data set and not used further in the analysis.

\subsection{Maximum Likelihood Imaging}
\label{sec:image}
A MLR test is used to determine the significance of flux detection at a single point in a given grid of source positions from the filtered data vector $\vec{o}$.  The MLR, $\lambda$, is calculated from the difference 

\begin{equation}
\lambda = C_{o} - C_{src}, 
\label{eqn:mlr}
\end{equation}

where $C_{o}$ and $C_{src}$ are the $\chi^{2}$ statistics for the null (no source is fitted to $\vec{o}$) and source (a source with free strength is fitted to $\vec{o}$) hypotheses respectively:

\begin{eqnarray}
C_{o} & = & \sum_{i}\frac{o^{2}_{i}}{\sigma_{i}^{2}} \\
C_{src} & = & \sum_{i} \frac{(o_{i} - \alpha e_{i})^{2}}{\sigma_{i}^{2}} 
\label{eqn:csrc}
\end{eqnarray}

The weighting factor, $\sigma_{i}$, is taken as the statistical error on $o_{i}$.  This becomes the average poisson error on the filtered BATSE count rate.  Assuming that the error on the modelled background is small;

\begin{equation}
%\sigma_{i}^{2} = \frac{\sum^{j=i + f}_{j=i-f}C{j}}{2f+1}.
\sigma_{i}^{2} = \frac{o_{i}}{2f+1}
\label{eqn:err}
\end{equation}

The response vector, $\vec{e}$, describes the expected position, shape and amplitude of the peaks as a function of source strength and position on the sky.  It is also differentiated so that the series of expected steps in the count rate for each source position are transformed into peaks, as shown in Fig.~\ref{fig:occresp}.  The differentiated response scales between 0 (no change in transmission) to $\pm 1$ (maximum change in transmission) as a function of the angular response of each LAD.  The scaling factor, $\alpha$, is proportional to the flux received from each point in the sky for all eight LADs,

\begin{equation}
\alpha = \left (\sum_{i}\frac{o_{i}e_{i}}{\sigma_{i}^{2}}\right )\left (\sum_{i}\frac{e_{i}^{2}}{\sigma_{i}^{2}}\right )^{-1}
\label{eqn:alpha}
\end{equation}

and has a statistical uncertainty given by,

\begin{equation}
\Delta\alpha = \left ( \sum_{i} \frac{e_{i}^{2}}{\sigma_{i}^{2}} \right ) ^{-1/2}
\label{eqn:deltaalpha}
\end{equation}

The distribution of $\lambda$ is the same as $\chi^{2}_{\nu}$ with $\nu$ degrees of freedom.  When searching for a known source $\nu = 1$ and hence the detection significance, in gaussian $\sigma$, is given by $\surd\lambda$.  Because of this relationship $\chi^{2}$ gives source significance directly.  Since $\alpha$ satisfies the condition of minimum $\chi^{2}$ it can be considered as a property of the data and the response vector and it is not a free fit parameter.  The test is easily extended to searches for new sources of gamma-rays at unknown positions; two extra degrees of freedom are added for the $(x,y)$ sky position and so $\nu = 3$ (\citet{jurgen}).

The likelihood image, $\lambda(x,y)$, is built up from the superposition of the arcs caused by the occultation of point sources by the Earth's limbs.  The effect of the occultation angle, $\beta$, on the angular resolution of the imaging technique was discussed in Sect.~\ref{sec:eot}.  A second effect improves the resolution at moderate $\beta$ when more than one occultation arcs are considered.  At $\beta\sim$45$^{\circ}$ the rising and setting limbs are almost orthogonal, and give good positional information in both the $x$ and $y$ directions of the map simultaneously.  In addition the natural precession of the CGRO orbit means that a number of angles are sampled every 53 days.

\subsection{LAD Response Matrix}
The calculation of response matrices to describe the sensitivity of
gamma-ray detectors as photon energy and incident angle changes is a
difficult task.  The approach used here is based on work by
\citet{batse_eot}.  The standard occulation technique was used to
observe the Crab nebula over a considerable portion of the CGRO
lifetime.  During this time the Crab was seen by all eight detectors
at various angles incident to their optic axes, $\theta$.  A function
was then fitted to the points for each LAD and energy channel assuming
azimuthal symmetry of the instrumental response.  The response, $R$,
is described by an equation containing 4 parameters, $A_{1}, A_{2},
A_{3} $ and $A_{4}$, which are unique for each energy channel and detector.  The equation for the response, taken from \citet{batse_eot}, is as follows:

\begin{equation}
R = A_{1}\cos(\theta)\exp^{\left(\frac{-A_{2}}{\cos(\theta)} + \frac{A_{3}}{\sin(\theta+A_{4})}\right)}
\label{eqn:crabcnts}
\end{equation}

This calibration gives maps of $\alpha$ in units of the normalised Crab Nebula response.

\begin{figure}
\resizebox{\hsize}{!}{\includegraphics{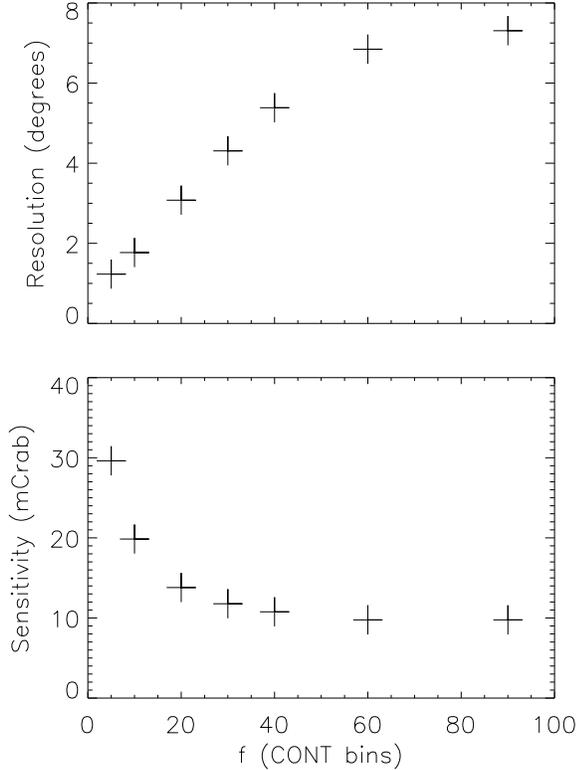}}
\caption{The flux sensitivity ($3 \sigma$) and angular resolution (FWHM) of the LIMBO technique for varying filter sizes, $f$, measured in number of 2.048 second CONT bins.  The measurements are based on one days observation of the Crab Nebula imaged with the LIMBO technique over a 0.1$^{\circ}$ grid of positions.}
%\vspace{150mm}
\label{fig:diffltscrab}
%\vspace{150mm}
\end{figure}
\subsection{Combination of Images}
\label{sec:addimages}
The sensitivity of the imaging technique can be
  increased, by approximately the square root of exposure time, by
  considering longer observations.  It is possible to run the maximum
  likelihood imaging technique, of Sect.~\ref{sec:image}, on any
  length of time required. However, in practice, the maximum length of time
  that can be considered in this way is limited to a few tens of days
  by the availability of computing resources, since large amounts of
  data need to be read into memory at once.  Also, once the image has been
  made, it is not possible to simply extract an image from a subset of
  the data without running the analysis on that data alone.  Hence it is attractive to be able to make images of short periods of time (e.g. daily) and combine them into longer, more sensitive observations at a later date.  This would also allow the time variability of bright sources, found in the high sensitivity image, to be investigated on shorter timescales.  

It is not possible to add individually produced likelihood images
together directly. This is seen from the dependence of
Eq.~\ref{eqn:csrc} on Eq.~\ref{eqn:alpha}, since $\alpha$ is
the ratio of two sums, formed from the complete data set in question, namely;  
\begin{equation}
F = \sum_{i}\frac{o_{i}e_{i}}{\sigma_{i}^{2}}
\label{eqn:F}
\end{equation}
\begin{equation}
D = \sum_{i}\frac{e_{i}^{2}}{\sigma_{i}^{2}}
\label{eqn:D}
\end{equation}
It is possible, however, to calculate the total flux value, $\alpha_{T}$, for
each image pixel by summing the values of $F$ and $D$ from shorter
time periods.  In this case $d$ will be used to index a number of
images, each made from single, day long, observations.  It is
important to emphasise,  since even for a constant source of gamma-rays, the values of $D_{d}$ may be different from image to image (i.e. due to data gaps, changing orientation of the spacecraft, etc.), that
\begin{equation}
\alpha_{T} =\left (\sum_{d}F_{d}\right)\left (\sum_{d}D_{d}\right )^{-1} \neq \sum_{d} \alpha_{d}
\end{equation}

Hence, if a data set, indexed by $k = \{1,\ldots,d,d+1,\ldots,N\}$, is split into two sets, with $C_{src_{a}}$ and $\alpha_{a}$ evaluated for $k = \{1,\ldots,d\}$, and  $C_{src_{b}}$ and $\alpha_{b}$ evaluated for $k = \{d+1,\ldots,N\}$, then it is clear that $C_{src_{a}} + C_{src_{b}} \neq C_{src_{T}}$ because
\begin{eqnarray}
\sum_{i=i(k=1)}^{i(k=d)}\frac{(o_{i} - \alpha_{a}
  e_{i})^{2}}{\sigma_{i}^{2}} &+&  \sum_{i=i(k=d+1)}^{i(N)}\frac{(o_{i} -
  \alpha_{b} e_{i})^{2}}{\sigma_{i}^{2}} \nonumber \\ & \neq & \sum_{i=i(k=1)}^{i(N)}\frac{(o_{i} - \alpha_{T}
  e_{i})^{2}}{\sigma_{i}^{2}}
\end{eqnarray}

Images of statistical significance of gamma-ray emission can be made
by scaling $\alpha_{T}$ by its statistical uncertainty,
$\Delta\alpha$, given by Eq.~\ref{eqn:deltaalpha}.  With this method it is also possible to combine images made from
individual BATSE energy channels, which allows the sensitivity of the
imaging technique to be increased whilst retaining spectral, as well
as temporal, resolution.

\subsection{Differential Filter Choice}
\label{sec:filt}
The choice of differential filter width, $f$, has implications for the sensitivity and angular resolution of the imaging technique.  Choosing a larger $f$ will give a more significant occultation peak and improve sensitivity, smaller $f$ returns less significant but sharper peaks that will lead to better angular resolution.  Fig.~\ref{fig:diffltscrab} shows the performance of the imaging technique for various choices of $f$.  The figure is based on a 53 day observation of the Crab Nebula using one BATSE energy channel ($\sim 25 - 35$ keV) and an image grid resolution of 0.1$^{\circ}$ .  The sensitivity improves roughly as $1/\surd f$ but is seen to level out as $f$ increases to the point where the timescale of variations in the background approach $f\times 2.048$ seconds.  For all choices of $f$ the peak flux in the map was located within $< 12\arcmin $ of the actual Crab position of $(l,b) = (184.56$$^{\circ}$ $, -5.78$$^{\circ}$ $)$.  It can be seen that $f \approx 10 - 20$ offers a good compromise between sensitivity and resolution.  It is clear however that different $f$ may be used to optimise the technique for particular science goals.
%; for example, new sources detected in an initial high sensitivity image could be isolated and located to by ``zooming in'' on the area of interest and using $f \approx 5$.

\begin{figure}[t]
\resizebox{\hsize}{!}{\includegraphics{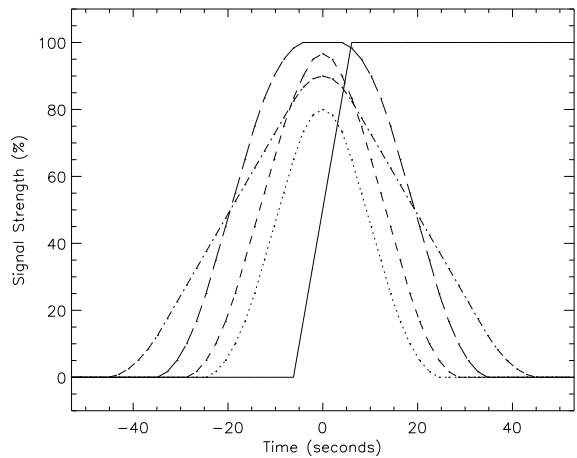}}
\caption{A  simple model to  show the  effect of  various differential filters on  an occultation step  of duration 12 seconds  (solid line).  Filters with  $f = 10$ were applied  with inner gaps, $f_{a}$, of  0 (dotted), 2 (short dash) and 5  (long dash) as well as a filter  with $f = 20$ and inner gap = 0 (dot-dash).}\label{fig:difflts}
%\vspace{150mm}
\end{figure}

It is possible to make an improvement to the sensitivity of a
particular chosen $f$ whilst retaining reasonable angular resolution.
The size of a differentiated occulation peak depends on the difference
between the mean counts calculated before and after the time bin in
question.  Since the occultation steps are not infinitesimal (they
take about 10 seconds), the mean
counts contain a contribution from the transition of the source from
full transmission to full occultation that acts to reduce the average difference between the two means.  This causes a reduction in sensitivity which becomes more important as $f$ decreases towards the step time of about 5-6 bins.  This can be avoided by adding an inner bound to the differential filter.  By making a small gap in the centre of the filter systematic errors in calculating the mean counts either side of a step are reduced but angular resolution can be retained.  Fig.~\ref{fig:difflts} and Table~\ref{tab:difflts} shows the effect of applying several different filters to a simple model of an occultation step.  Clearly $f = 10$ with a small inner gap offers the best combination of angular resolution (given by the HWHM of the curves) and sensitivity.

\begin{table}[h]
  \begin{center}   \caption{Approximate    sensitivity   and   angular
resolution of differential  filters used in Fig.~\ref{fig:difflts}.}
\renewcommand{\arraystretch}{1.2}
\begin{tabular}[h]{cccc} \hline $f$& $f_{i}$ & Sensitivity & Angular
    resolution \\ (bins) & (bins) & (\%) & ($^{\circ}$) \\ \hline 10 &
    0 & 80 & 0.75 \\ 10 & 2 & 97 &0.89\\ 10 & 5 & 100 &1.21\\ 20 & 0 &
    90 &1.38\\ \hline \\
\end{tabular} 
\label{tab:difflts} 
\end{center}
%\vspace{150mm}
\end{table}

A possible extension  to the differential filtering technique may be to  use different  resolution maps to  build up images  of extended emission.   By choosing  different  values of  $f$  that optimise  the coverage of  spatial frequencies  in the image  it may be  possible to investigate, for example, diffuse gamma-ray emission from the Galactic Plane \citep{mattsthesis}.

\subsection{Maps of statistical significance of source emission}
\begin{figure*}
\centering
\includegraphics[width=17cm]{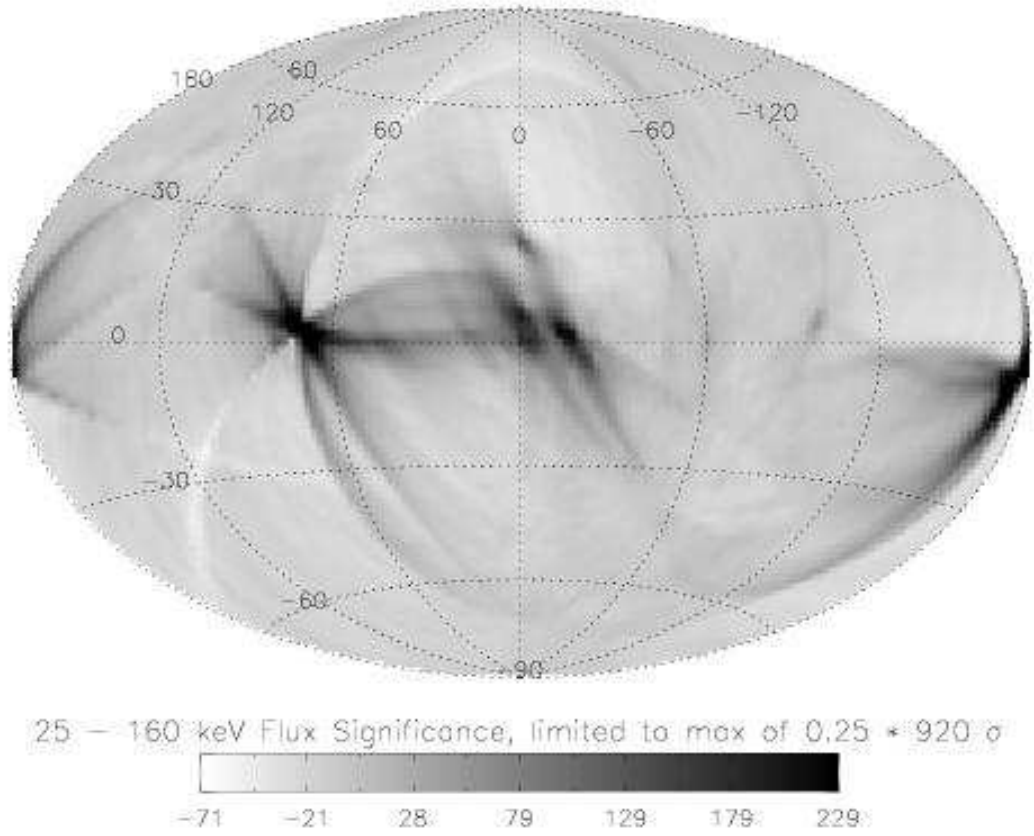}
\caption{Map of the significance of 25 - 160 keV flux from the whole sky for a total of 489 days from TJD 09448 to TJD 09936 made using LIMBO.  The colour scale of the map has been limited to a quarter of the peak in order highlight details at lower significance.  Many well known, bright sources are visible and the peak significance of 920 $\sigma$ is consistent with the position of the Crab Nebula.  There is a limited systematic across the whole image but there are serious artefacts which limit the usefulness of the image for observing weak sources or crowded regions.}\label{fig:rawmap}
%\vspace{150mm}
\end{figure*}

An all-sky map for 489 days in the period TJD 09448 - 09936, produced from 7 CONT energy channels (25 - 160 keV), is shown in Fig.~\ref{fig:rawmap}.  Using simple statistical arguments, and neglecting any systematic background, the map has a $3\sigma$ sensitivity of 3 mCrab.  This can be considered as the first all-sky gamma-ray image to be produced since the HEAO1-A4 mission.  Although it is easy to see the brightest sources in the sky, such as the Crab Nebula and Cygnus X-1 it is clear that the imaging technique is heavily affected by artefacts stemming from the bright sources.  This is a particular problem in the crowded region of the Galactic Centre, where source confusion is a problem.  

The artefacts seen in the figure are images of the Earth's limb from individual occultations.  An individual  limb implies a probability of a  source  being  observed  anywhere  along its  length.   The  source location accuracy improves as more limbs are superimposed and the size of the  likelihood peak increases.  Although the  brighter sources can be located  accurately from their likelihood peaks  the artefacts from limbs passing through the source position may be more significant than the likelihood peaks found at the position of fainter sources.

\section{Image Artefact Removal I: Removal of Individual Sources}

\begin{figure*}
\centering
\includegraphics[width=17cm]{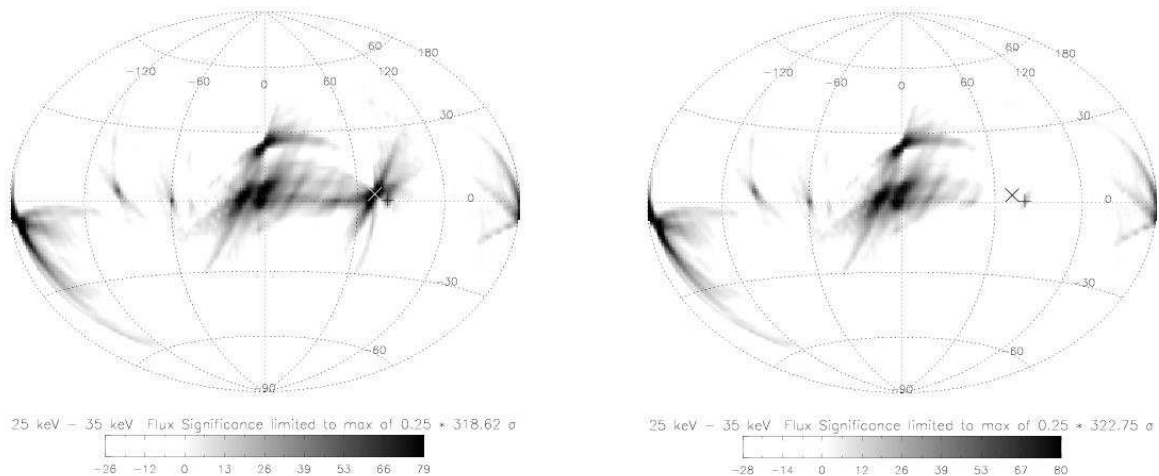} 
\caption{An example of the removal of source artefacts by making a second pass of the data with the likelihood function adjusted to include an inverse of the response at the position of Cygnus X-1 (maked by `$\times$').  The left image shows the LIMBO all-sky map for data from TJD 9450 - 9900, 25 - 35 keV.  In the right image Cygnus X-1 has been removed.  Note the faint response near Cygnus X-1 that is apparent in the right image, which is consistent with the position of Cygnus X-3 (marked by `$+$').}\label{fig:nocygx1}
%\vspace{150mm}
\end{figure*}

As mentioned earlier the diameter of the Earth seen from CGRO is about 140$^{\circ}$.  Hence the limbs crossing through a source position can have an effect over a large region  of the sky.  This is especially obvious from the map of Fig.~\ref{fig:rawmap} for the  bright sources, such as the Crab and Cygnus X-1.   The effects of  these sources can  be removed from the data by making two passes of the likelihood calculation.  By first calculating the response function for a particular source position, the effect on the entire image of a source at this position can then be removed by adding the inverse of the response to  the likelihood calculation. Fig.~\ref{fig:nocygx1} shows the removal of Cygnus X-1 and all its associated artefacts from an image. When the confusing influence of Cygnus X-1 has been removed it is possible to see a faint source that is consistent with the position of Cygnus X-3.

The iterative removal of individual sources in this manner may not, however, be an ideal method of reducing the influence of image artefacts so that fainter sources may be observed:  The response measured from each source position will include a systematic effect due to artefacts from sources at other positions.  This will lead to a cumulative error in estimating the flux each time a source is removed, although the seriousness of this effect may be limited if only the most significant of sources are removed first.  Finally repeated re-running of the LIMBO code will clearly incurr overheads in computing time that may be prohibitive.  

\section{Image Artefact Removal II:  Image Cleaning Algorithm}
All pictures or maps produced by any imaging technique consist of the
true spatial distribution of light (the object function) convolved
with some distorting function specific to the instrument or technique
being used (the impulse response).  In optical astronomy, the impulse
response is minimised by carefully designing the instrument to reduce
aberrations leading to a compact, symmetrical Point Spread Function
(PSF).  At other wavelengths, where instrument designers have less
control over the optical design, a post observation approach must be
applied to reduce distortions in the image.  One such example is in radio interferometry where the CLEAN algorithm \citep{hogbom74} has been used for many years to remove artefacts from radio maps introduced due to the finite number of radio dishes, the distribution of the baselines used and the motion of the Earth relative to the source \citep{tan86}.  

The CLEAN algorithm works by defining two data spaces; the raw map $M_{R}$ (consisting initially of the original sky image) and the clean map $M_{C}$ (of the same dimensions as the raw map, but initially empty).  The algorithm proceeds by first finding the brightest pixel in $M_{R}$ and generating the expected PSF for this position.  A small fraction of the PSF, defined by a gain set by the user, is removed from $M_{R}$.  The total flux that has been removed is integrated over the PSF.  If required, this flux can than be convolved with an ideal PSF, such as a Gaussian distribution with a realistic width, and added to $M_{C}$ at a position centred on the position of the brightest pixel in $M_{R}$.  This process is repeated iteratively until some predefined criterea, such as the minimum significance of the brightest pixel remaining in $M_{R}$, is reached.  At this point what remains of $M_{R}$ should resemble intrinsic noise, which can be added to $M_{C}$ to produce a more realistic final image.

At any time the shape, size and orientation of a BATSE occultation limb can
be calculated given information about the orbit of CGRO and the detector
pointings. Since the artefacts produced by point sources in LIMBO are
due to the accumulation of occultation limbs, an impulse response
function may be calculated for any position in the sky based on the
knowledge of the history of the position of the Earth's limb. This
response to a point source is conceptually similar to a PSF for the
observations, although it should be highlighted that the physical
interpretation differs from the PSF used in other imaging modalities.
For example, in optical astronomy the PSF represents the characteristic
way the imaging system redistributes the point source of flux in the
image; a point source of photons in the sky becomes extended in the image. In LIMBO the PSF represents the characteristic
way the imaging system redistributes the probability of the estimated
position of the source; a point source in the sky produces an image that
is consistent with emission from a number of different, but mutually
exclusive, positions.  

The PSF for any particular sky position is highly complex, as is shown in the example in Fig.~\ref{fig:crabpsf50}.  However, if the PSF of a selected source position can be calculated it is possible to subtract that source's contribution to the total image.

\subsection{Calculation of the LIMBO Point Spread Function}
The PSF for the LIMBO technique is highly complex as it is defined by the set of loci of the Earth's limb for any given source position and depends on the orientation of the detectors with respect to the position of the spacecraft in the orbital precession cycle.  Consequently the PSF for any given source is temporally, spatially and spectrally variant.  It is possible to make an accurate determination of the PSF by creating a dataset from the known response vector of a target source and applying the LIMBO code to it.  However, this is an extremely inefficient approach as in an all-sky image the majority of pixels for which LIMBO generates a response and evalutes the MLR never undergo an occultation simultaneously with the target source.  A more efficient approach has been developed by \citet{mattsthesis}, which uses knowledge of the position of the Earth's limb over time to restrict the number of pixels LIMBO tests for a response from.  For each time bin during the occultation, the pixels through which the Earth's limb passes can be identified by using the spherical polar coordinates of the Earth's centre (as seen from CGRO), which are recorded along with the counts, in the BATSE data stream.  Once all of the relevant pixels have been identified LIMBO can be run on this smaller set.  For a 2$^{\circ}$ resolution grid of sky positions this approach leads to a reduction in the number of pixels for which the MLR is calculated from 16,471 to $\sim 1000$, which allows a significant reduction in the amount of computing time required.

\subsection{Choosing a Linear Cleaning Parameter}
An essential requirement for the CLEAN algorithm to work is that the imaging property being reduced is linear, i.e. the quality of the final map should not be a function of the order in which contributions to the background from individual sources were subtracted.  This is not true for the MLR, which is based on a geometrical sum.  However, $F$, (recall Eq.~\ref{eqn:F}) can be shown to be linear for an image containing a superposition of $S$ sources.  The differential filter of Eq.~\ref{eqn:difffilt} can be written in a simplified form as

\begin{equation}
\label{eqn:filtsimp}
o_{i} = \frac{1}{f + 1}\sum_{j(i)}r_{j}
\end{equation}

If the filter is associated with an operator, $O$, such that $O(r_{j})
= o_{i}$, and $r_{j} = \sum^{s=S}_{s=1}r_{sj}$ is substituted into Eq.~\ref{eqn:filtsimp} it can be shown, since $r$ is linear, that

\begin{equation}
o_{i} = \frac{1}{f+1}\sum_{s}\sum_{j(i)}r_{sj} = {\sum_{s}O(r_{sj})}
\end{equation}

Hence the law of superposition, $O(\sum_{s=1}^{s=S}r_{sj}) = \sum_{s=1}^{s=S}O(r_{sj})$, is true and it has been shown that the differential filter is a linear operator.  Since both $e_{i}$ and $\sigma_{i}$ are constants it follows that $F$ is also linear,

\begin{equation}
F = \sum_{s}\sum_{j(i)}\frac{O(r_{sj})e_{i}}{\sigma_{i}^{2}}
\end{equation}

It does not follow that the flux maps are linear.  However, since $D$ (Eq.~\ref{eqn:D}), is a constant it is possible to run the cleaning algorithm on $F$ and use the results of this to derive the cleaned flux and log-likelihood maps. 

\subsection{Identifying which Pixel to Clean}
For some systems the identification of the sky pixel location at which to clean, $p_{(l,b)}$, is trivial - it is simply the brightest pixel.  For the LIMBO system this is not necessarily the case since the peak in $F$ is not necessarily in a pixel consistent with the true source position leading to calculating the PSF for an incorrect position.  For this reason $p_{(l,b)}$ is found by first generating the PSF for the most significant pixel and then cross-correlating this on a grid surrounding the most significant pixel.  The pixel position with the highest correlation coefficient is then assumed to be the correct pixel.  A second PSF is regenerated for this position and a small fraction of it subtracted from the image and added to the clean map.

Choosing the incorrect $p_{(l,b)}$ is particularly problematic for the LIMBO images since, for these initial tests of the technique, the pixels are 2\degr$\times$ 2\degr and hence the error can be great. Even if the correct pixel is chosen the PSF will still not be generated at exactly the correct position since it must be generated at the centre of the pixel; the maximum discrepancy being $\sqrt{2^{2} + 2^{2}} = 2.8^{\circ}$. In order to address this problem two strategies are introduced. Firstly there is a pre-cleaning stage where the majority of the contributions from the brightest known sources, with well defined positions, are removed. Secondly, a catalogue of probable sources is checked as each $p_{(l,b)}$ is found.  The pre-cleaning stage is particularly important since the systematic error introduced into the final $M_{C}$ map by cleaning with the incorrect PSF is dependent on the total amount of flux removed with that PSF. Hence, if bright sources, such as the Crab, are cleaned with only a slightly incorrect position, the resulting errors will be large compared to the faint sources whose detection is the ultimate goal.

 For a list of known sources, the PSF at the exact position each of the sources is generated and used prior to running the cleaning algorithm to remove most (typically 95\%) of the contribution to $M_{R}$. As new sources are discovered they may be added to this list and the cleaning algorithm run again to find the next level of sources.  As the cleaning algorithm progresses each $p_{(l,b)}$ CLEAN pixel is also compared to a list of known sources that it is believed, though not definitively known, to be hard X-ray sources. This list comprises, amongst others, the HEAO catalogues, known AGN, and transient Galactic sources. If within $p_{(l,b)}$ there is a candidate source then the PSF may be generated at the exact position of this source. Care must be taken however in crowded regions.

\section{LIMBO Image Cleaning - Preliminary Results}
\begin{figure*}
\resizebox{\hsize}{!}{\includegraphics{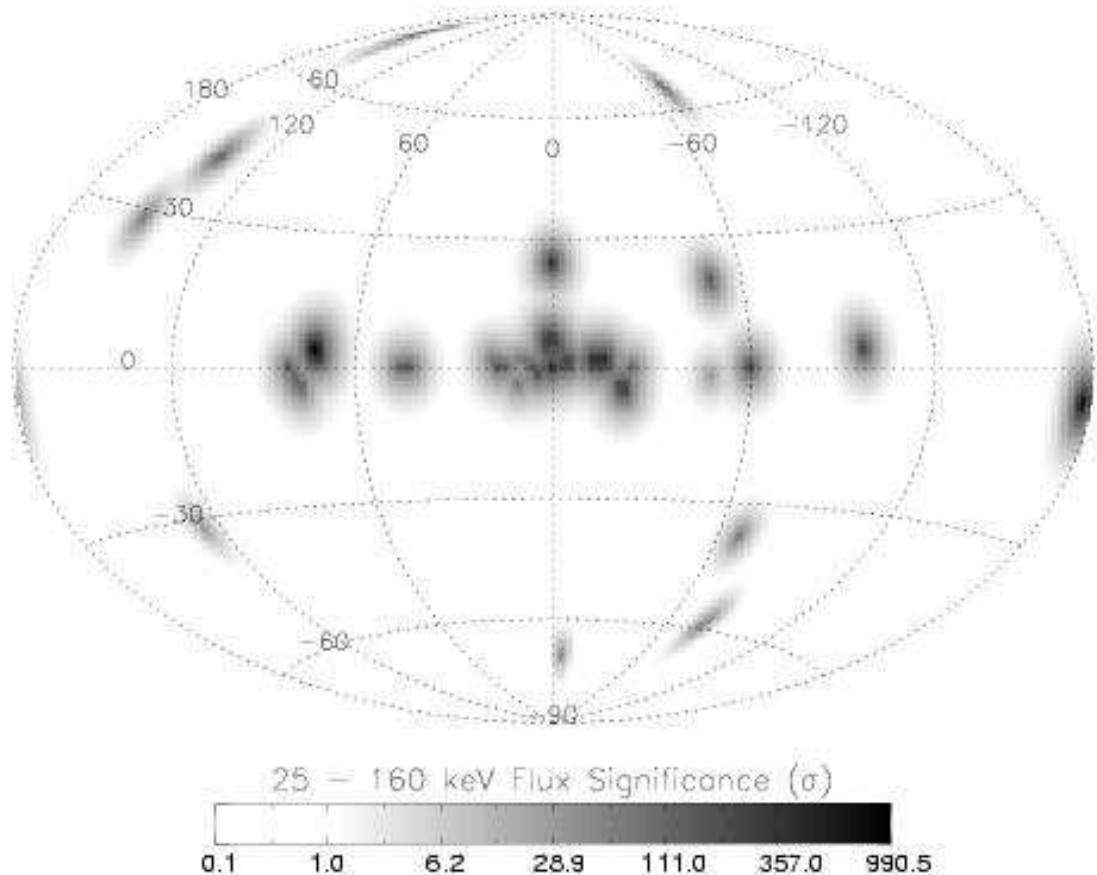}}
\caption{The results of performing the image cleaning technique on the all-sky map of Fig.~\ref{fig:rawmap}.  See Tables~\ref{tab:cat} and \ref{tab:cat2} for a summary of the identified sources from the image.}\label{fig:clean}
%\vspace{150mm}
\end{figure*}

The results of applying the cleaning algorithm described above can be seen by comparison of Fig.~\ref{fig:clean} and Fig.~\ref{fig:rawmap}.  Sources were extracted from the map using the SExtractor 2.2.2 (SE) software,
developed
by \citet{sextractor}, with the following settings:
DETECT\_THRESH set to 1 sigma; CLEAN\_PARAM (cleaning efficency) of 1.0;
a wavelet pass-band filter, mexhat\_1.5\_5x5.conv.  The fluxes and
their associated errors are measured using SE's ISO\_CORR aperture option.

Two maps were produced to use with SE; one containing the cleaned components only and one with the residual flux, remaining after cleaning, added back to the cleaned components.  The latter map is of course the more scientifically realistic map and when running the SE code directly on it produced a catalogue of detections in the image, shown in Table~\ref{tab:cat}.  The fluxes of the sources in the table have been normalised to the Crab flux.  

The sources extracted by SE were compared against the INTEGRAL Science Data Centre (ISDC) catalogue of high energy sources \citep{ebisawapaper, ebisawacat}, which contains over 1000 sources with flux estimates in the 20 - 200 keV band.  Points of emission in the map have been identified with sources from the catalogue, based on their positions and estimated brightnesses.  It should be noted that, due to the inherent variability of gamma-ray sources, \citet{ebisawapaper} report large uncertainties in some of their flux estimates.  In addition, the catalogue and LIMBO spectral bands are not equal and so, in some cases, these source identifications are subjective and should be treated with caution.  Any source which lies within 2.8$^{\circ}$ of its catalogue position is within the diagonal size of one image pixel.  The angular resolution of the LIMBO images is limited by the 2$^{\circ}$ grid spacing.  Hence there are some cases where it is not possible to associate the detection with only one known source.  Source confusion is always likely to be a problem, especially in the Galactic Centre, with a technique such as Earth Occultation Imaging.  However, recall from Sect.~\ref{sec:eot} that it is possible to improve the angular resolution of the LIMBO technique to $0.5^{\circ}$.

\begin{table*}
\begin{center}   \caption{List of sources detected by BATSE with LIMBO in the 25 - 160 keV band for the  period spanning TJD 09448 - 09936.  The fluxes have been normalised to the Crab Nebula flux.}
\renewcommand{\arraystretch}{1.2}
\begin{tabular}[h]{clccc} \hline
Measured Position & Source Name & Known Position & Position Error&  BATSE Flux  \\
(l,b)             &             & (l,b)          & (degs)             &(mCrab)  \\ 
\hline
(71.9, 3.8)   & Cygnus X-1  & (71.34, 3.07)   &  0.9 & 1142  $\pm$ 18 \\

(-175.9, -5.9)& Crab        & (-175.44, -5.78)&  0.5 & 1000  $\pm$ 17 \\

(0.0, 6.9)   & GRO J1719-24 & (0.18, 7.02)   &  0.2  & 214  $\pm$ 9 \\
             & H 1705-250   & (-1.41, 9.06)  &  2.6  &   \\

(-12.0, 2.0) & 4U 1700-377  & (-12.2, 2.2)   &  0.3  & 209  $\pm$ 8 \\
             & GX 349+2     & (-10.90, 2.74) &  1.3  &           \\

(-2.0, 0.9)  & GRS 1734-292   & (-1.11, -1.41) & 1.0 & 201  $\pm$ 9 \\
             & H 1743-322   & (-2.87, -1.61) & 2.7   & \\

(0.0, 0.2)   & XTE J1748-288& (0.68, -0.22) &  0.8   & 180 $\pm$ 8 \\
             & 1E 1740.7-2942   & (-0.88, -0.11) & 0.9  & \\
             & GX 359+2  & (-0.43, 1.56)    &  1.4   &\\
	     & GX 3+1    & (2.29, 0.79)     &  2.4   &\\

(-16.0, 2.0) & OAO 1657-415  & (-15.64, 0.31)&  1.7 & 128 $\pm$ 7\\

(0.1, 24.0)  & Sco X-1    & (-0.90, 23.78)   &  0.9  & 114  $\pm$ 7 \\

(-21.5, -4.7)& GX 339-4   & (-21.06, -4.33)  &  0.6  & 93  $\pm$ 6 \\

(-96.3, 3.7) & Vela X-1   & (-96.94, 3.93)   &  0.7  & 83  $\pm$ 6 \\

(16.7, 1.4)  & GX 17+2    & (16.44, 1.28)    &  0.3  & 77  $\pm$ 8 \\

(122.0, -30.0) & Mkn 348 & ( 122.27, -30.91) & 0.9   & 51  $\pm$ 8 \\
          %    & 2A 0042+323 & (121.34, -29.83) & 0.6 & 1     & \\

(4.6, -1.4) & GX 5-1     & (5.07, -1.02)    &  0.6   & 46  $\pm$ 5 \\

(44.3, 0.0)  & GRS 1915+105 & (45.40, -0.23) &  1.1  & 43 $\pm$ 5 \\

(-2.00, 6.0) & GRO J1719-24 & (0.14, 6.99)   &  2.4  & 42  $\pm$ 5 \\

(-60.00, 0.0) & GX 301-2    & (-59.90, -0.04)&  0.1  & 42  $\pm$ 4 \\

(80.00, 0.6)  & Cygnus X-3  & (79.84, 0.69)  &  0.2  & 36  $\pm$ 5 \\

(-50.00, 19.5) & Centaurus A& (-50.48, 19.42)&  0.5  & 33  $\pm$ 5 \\

(-158.83, -4.0)& H 0614+091 & (-159.11, -3.38)&  0.7  & 31  $\pm$ 8 \\

(-24.00, 0.0) & H 1624-490 & (-25.08, -0.26)  &  1.1  & 17  $\pm$ 3 \\
%               & 4U 1630-47 & (-23.09, 0.26)  &  0.9 &   \\

(156.00, 76.0) & NGC 4151 & (155.08, 75.06) &  1.0    &16  $\pm$ 3 \\

(138.00, 42.0) & M 82     & (141.41, 40.57) &  2.9    &14  $\pm$ 3 \\

(-70.00, 66.0) & 3C 273   & (-70.05, 64.36) &  1.6    &12  $\pm$ 3 \\

%(-2.00, -70.0) & NGC 7582 & (-11.92, -65.70)  & 5.7  & 9  $\pm$ 3 \\

\hline
\end{tabular}
\label{tab:cat}
\end{center}
\end{table*}

\begin{table*}
  \begin{center}   \caption{List of additional sources visible in Fig.~\ref{fig:clean} but not included in Table~\ref{tab:cat}.  These sources were extracted using the locations where PSFs were generated in the LIMBO analysis as an input to the SExtractor package.}
\renewcommand{\arraystretch}{1.2}
\begin{tabular}[h]{clccc} \hline
Measured Position & Source Name & Known Position & Position Error & BATSE Flux  \\
(l,b)             &             & (l,b)          & (degs)             &(mCrab)\\ 
\hline
 (-79.0, -59.0) & ESO 198-24 & (-88.36, -57.95) & 5.0      &  62 $\pm$ 8 \\
 (-5.0, -70.0)  & NGC 7582   & (-11.92, -65.70) & 5.1      & 55 $\pm$ 8 \\
 (146.1, 28.0)  & Mkn 78     & (151.10, 29.78)  & 5.0      & 40 $\pm$ 6 \\
 (-68.0, -38.0) & 4U 0357-74 &(-71.42, -37.29)  & 2.8      & 38 $\pm$ 6 \\
 (-47.9, -2.0)  & Circinus Galaxy  &(-48.67, -3.81) & 2.0 & 28 $\pm$ 5 \\
 (76.8, -2.0)   & EXO 2030+375  & (77.15, -1.24) & 0.8    & 22 $\pm$ 5 \\
\hline
\end{tabular}
\label{tab:cat2}
\end{center}
\end{table*}

A number of sources visible in Fig.~\ref{fig:clean} were not, however, detected by SE.  Table~\ref{tab:cat2} shows a list of sources not included in Table~\ref{tab:cat}, which was compiled by using the positions where PSFs were generated during the cleaning process as a guide for SE to perform its source extraction from the cleaned~+~residuals flux map.  This method assumes that the cleaned detections are real.  It should be noted that fake sources, added at random positions in the input catalogue, will return essentially zero flux.  Hence guiding the source extraction in this manner does not produce false detections.  The measured fluxes of these sources indicate the possibility of sources at that position, but as they cannot be independently measured may not be entirely trusted.  It should be noted that the positional error for many of these source identifications is larger than for those in Table~\ref{tab:cat}.  In the Cygnus region this analysis results in the detection of EXO 2030-375 in addition to Cyg X-1 and Cyg X-3.  

\section{Discussion and Future Work}
The first true all-sky image of the gamma-ray sky since HEAO1-A4 has been produced, in a single 25 - 160 keV energy band, using the LIMBO maximum likelihood imaging technique with BATSE occultation data.  In order to produce the all-sky map it has been necessary to develop several new, novel techniques, such as: An accurate background model for the BATSE experiment (BAMM; \citet{bammpaper}); Methods for producing all-sky images from BATSE occultion data; A method for cleaning artefacts from maximum likelihood images.  The all-sky map is based on $\sim$ 500 days of data and shows the location of about 25 known gamma-ray sources, including several extragalactic objects at $4-5 \sigma$ level ($\sim$ 15 mCrab). 
 
In almost all cases, LIMBO has identified emission from within one pixel of a source from the ISDC high energy catalogue.  However, whether all the detections from the LIMBO map are real sources of gamma-ray emission, that have been associated with the correct object from the catalogue, can not be determined at present.

It is clear that the images presented here suffer from the coarse $2^{\circ}$ resolution of the imaging grid, which makes identification of sources in crowded regions, such as the Galactic Centre, very difficult.  The performance of LIMBO in these cases can be improved significantly by simply increasing the resolution of the imaging grid towards the theoretical imaging limit of $0.5^{\circ}$, although this would be at the expense of an increase in computing time.

Since a single energy band has been considered, no measurements have been made of the spectra of any of the detected sources and fluxes are given only in comparison to the detected Crab Nebula flux.  The all-sky map was made by summing the images from 7 CONT energy channels.  Tests have shown that there is significant imaging response in the individual energy channels and further information could be gained from summing energy channels above 160 keV.

It is envisaged that further investigation will be made of the spectral performance of LIMBO with the aim of calibrating the response so that the continuum spectra of gamma-ray sources may be examined.  A rough estimate of the statistical sensitivity of the LIMBO technique to persisitent gamma-ray sources can be found by scaling the detection significance of the Crab nebula (990$\sigma$) in the 500 day image.  This gives a 5$\sigma$ sensitivity in the 25-160 keV range of $\sim$5 mCrab.  If the whole BATSE CONT dataset of $\sim$3300 days is considered then the sensitivity is improved by the square root of the ratio of the observation times, leading to a sensitivity of $\sim$2 mCrab.  Depending on the brightness of the source, it should also be possible to make time resolved spectra for the duration of the BATSE mission.  Assuming an equal contribution to the integrated flux from each of the 7 CONT energy channels and using the same simple statistical arguments as above, it should be possible to make broad spectra for gamma-ray sources with a constant flux of $\sim$6 mCrab.  Work continues to produce a publically accessible database of daily all-sky maps at $0.5^{\circ}$ resolution for several energy channels of the full 9 years of available BATSE data. 

\begin{acknowledgements}
The authors would like to thank J. Kn{\"o}dlseder for providing the initial concept of maximum likelihood imaging with the BATSE Earth Occultation technique.  G.J. Fishman, A. Harmon, C.A. Wilson and others at the National Space Science and Technology Center, Huntsville are also thanked for their considerable assistance in providing data, gamma-ray source measurements and for their helpful comments during the preparation of this work.
\end{acknowledgements}

\bibliographystyle{aa}
\bibliography{shaw0026}

\end{document}